\documentclass[twocolumn,showpacs,amsmath,amssymb,showkeys]{revtex4}

\usepackage{graphicx}
\usepackage{dcolumn}
\usepackage{bm}
\usepackage{rotating}
\usepackage{verbatim}

\makeatletter 
\gdef\@ptsize{0}
\let\@currsize\normalsize 
\makeatother 
\usepackage{setspace} 
\doublespace 

\usepackage{xcolor}

\begin{document}

\title{Propagation of Economic Shocks in Input-Output Networks:\\A Cross-Country Analysis\footnote{Thanks to Nobuyuki Hanaki and Alan Kirman for stimulating comments.}}


\author{Martha G. Alatriste Contreras}
\affiliation{Aix-Marseille School of Economics, CNRS, EHESS \\ 2 rue de la Charite, 13236 Marseille cedex 02 \\  Email: \texttt{martha.alatriste@ehess.fr}}%

\author{Giorgio Fagiolo}%
\affiliation{Institute of Economics, Scuola Superiore Sant'Anna\\ Piazza Martiri della Libert\`{a} 33, 56127 Pisa (Italy)\\ Email: \texttt{giorgio.fagiolo@sssup.it}}


\begin{abstract}
\noindent \textbf{Abstract. }This paper investigates how economic shocks propagate and amplify through the input-output network connecting industrial sectors in developed economies. We study alternative models  of diffusion on networks and we calibrate them using input-output data on real-world inter-sectoral dependencies for several European countries before the Great Depression. We show that the impact of economic shocks strongly depends on the nature of the shock and country size. Shocks that impact on final demand without changing production and the technological relationships between sectors have on average a large but very homogeneous impact on the economy. Conversely, when shocks change also the magnitudes of input-output across-sector interdependencies (and possibly sector production), the economy is subject to predominantly large but more heterogeneous avalanche sizes. In this case, we also find that: (i) the more a sector is globally central in the country network, the largest its impact; (ii) the largest European countries, such as those constituting the core of the European Union's economy, typically experience the largest avalanches, signaling their intrinsic higher vulnerability to economic shocks.

\end{abstract}





\maketitle

\section{Introduction\label{Sec:Introduction}}
Studying the mechanisms through which shocks diffuse in economic systems is of a foremost importance to devise predictive tools and policy measures that can help regulators to dampen aggregate fluctuations and reduce the likelihood of systemic crises \cite{Haldane_2009,Haldane_2011}. Indeed, as the recent financial and economic crisis has clearly shown, shocks can quickly percolate among countries and through their industrial sectors, turning country-specific shocks originated in the financial sectors into worldwide recessions hitting the real side of the economy as well \cite{BIS_2011}.

Whereas a huge literature has analyzed the mechanisms of contagion in banking and financial networks \cite{Upper_2011,Chinazzi_Fagiolo_2013}, much less is known about how the network structure of interdependencies between the sectors of an economy shapes the process of diffusion of exogenous shocks. From an empirical perspective, a handful of studies have characterized the structure of input-output (IO) networks to better understand the topology of inter-sectoral dependencies and their repercussions at the macroeconomic level \cite{Slater1978,Blochl2011,McNerney2012}. Conversely, from a theoretical perspective, a few studies have highlighted that the topology of IO linkages \cite{Leontief1936} between economic sectors can amplify small productivity shocks and generate recessions \cite{Acemoglu2012,Acemoglu2013}. However, these works have mostly employed very abstract and often unrealistic IO network structures, very different from what one observes in empirical studies. Therefore, shock propagation and cascade formation in IO network structures mimicking the real-world structure of industrial interlinkages is still poorly understood \cite{Xu2011}.

This work begins to fill this gap by blending together economically-meaningful shock-diffusion models and Eurostat data on IO tables \cite{Dietzenbacher1998} for European Union (EU) countries in year 2005. For each country, we employ IO tables to build a weighted-directed IO network describing the structure of dependencies between sectors. We employ the observed IO networks to calibrate and simulate a number of network-diffusion models. 

In the last years, shock propagation in economic networks has been mostly explored using models borrowed from the literature studying propagation of infectious diseases. Examples include applications to shock diffusion in financial or trade network using Susceptible-Infected-Susceptible (SIS) \cite{Kucuk2012} or Susceptible-Infected-Recovered (SIR) diffusion models \cite{Toivanen2013,Kucuk2012}. Here, we depart from such literature to study simple but economically-meaningful diffusion models that differ as to their assumptions about where a shock comes from, and how it is locally diffused in the economy \cite{Lee2011,Karimi2013}. In particular, we explore three basic models, exploring two main dimensions: (i) the origin of the shock and its impact on IO linkages; (ii) the possibility that after a shock hits a sector, also its production levels are accordingly adjusted. By doing so, we try to disentangle the roles played by the nature of the shock, the structure of sectoral interdependences, and the dynamics of production during propagation. We then explore, for each different model, the likelihood that shocks generate avalanches (e.g. cascades) in the system and the distribution of their extent. Furthermore, we ask whether these outcomes depend on the type of shock (e.g., impacting final demand vs affecting country economic capacities and technological interlinkages), the size of the economy, the sector where the shock has originated, and the topological properties of the underlying IO networks.  

Our main result is that pre-crisis impact of economic shocks on European countries strongly depends on the nature of the shock and country size. Shocks that impact on final demand without changing, during diffusion, production and the technological relationships between sectors have on average a larger but very homogeneous impact on the economy. Instead, shocks that can change input-output interdependencies (and possibly sector production) as they percolate through the economy engender predominantly large but more heterogeneous avalanche sizes. Typically, the more a sector is globally central in a country IO network, the larger its impact on the economy when it is hit by a shock. We also find that countries constituting the core of the European Union typically experience the largest avalanches. This signals their intrinsic higher vulnerability to economic shocks.

The rest of the paper is organized as follows. Section \ref{Sec:Data_Methods} briefly describes the models and the data employed to calibrate them \footnote{See the Supplemental Material [SM] section for more technical details related to data and methods.}. Results are reported in Section \ref{Sec:Results}. Finally, Section \ref{Sec:Conclusions} concludes.

\section{Data and Methods\label{Sec:Data_Methods}}
Consider a closed economy composed of $S$ industrial sectors linked via a set of input-output relations described as a weighted directed graph with self loops. A node in the graph is a sector and a weighted directed edge represents an economic transaction conducted between sectors to buy or sell inputs used in the production process \cite{Blind1974,Amaral2007,Blochl2011}. The weighted adjacency matrix $Z$ has entries $z_{ij}>0$, $i\neq j$, proportional to value of the inter-sectoral flow from sector $i$ to sector $j$, i.e. the output of sector $i$ to be used as input in sector $j$'s production process. If a flow between $i$ and $j$ is zero, then the two sectors are not connected. Strictly-positive self loops $z_{ii}>0$ capture the idea of a sector using its own product as input. Since the network is directed, in general we have $z_{ij} \neq z_{ji}$. 

\subsection{Diffusion Models\label{SubSec:Models}}
We use IO networks as the backbone over which shocks are possibly transmitted from a sector to a neighboring one along the production chain. We are interested in understanding how shocks initially originating in a certain area of the technological space of a given country can possibly percolate across the entire structure of the economic system, i.e. how local shocks can have global effects \cite{Acemoglu2012}. We employ three different shock-diffusion models, which we explain in what follows. We focus on \textit{progressive} diffusion processes \cite{Kleinberg2007}, where once a sector has been hit by a shock it cannot be hit again \footnote{A \textit{non-progressive} process is instead one where a sector may be hit many times, as it happens with a person who recovers after having caught a disease and then becomes susceptible of contagion again, instead of becoming immune.}.

\subsubsection*{Model 1}

The first diffusion framework we study is a standard input-output model \cite{Leontief1936} where shocks come from final demand. In the model, sectoral output linearly depends on the input requirements from all sectors in the economy, and final demand from households, government, exports, and capital investment:

\begin{equation}\label{eq:IOM1}
\mathbf{x = Z1+d}
\end{equation}
where $\mathbf{x}$ is the $S \times 1$ output vector, $\mathbf{Z}$ is the inter-sectoral input-output matrix defined above, $\mathbf{1}$ is a column vector of ones, and $\mathbf{d}$ is the $S \times 1$ column vector of final demand. Simple algebra [SM] allows one to get sectoral production as a function of final demand and the matrix of technical coefficients $\mathbf{\Theta}=\{ \theta_{ij}\}=\{z_{ij}/x_{j}\}$:

\begin{equation}\label{eq:IOM3}
\mathbf{x = (I - \Theta)^{-1} d = Ld}
\end{equation}
where $\mathbf{L}=\mathbf{(I-\Theta)}^{-1} = [l_{ij}]$ is an $S \times S$ matrix known as the Leontief inverse or the total requirements matrix.

The consequence on output of shocks hitting the final-demand vector $\mathbf{d}$ can be easily modeled. Suppose that final demand of sector $s$ is hit by a shock that results in new levels equal to $d_s+\epsilon_s$. Then the ensuing change in the output vector reads: 
\begin{equation}\label{eq:IOM6}
\mathbf{\Delta x = L\epsilon},
\end{equation}
where $\mathbf{\epsilon}=(\epsilon_1,\dots,\epsilon_S)$. This implies that the impact of a single negative final-demand shock of size $f>0$ originated from sector $s$ will be equal to $-f\mathbf{L^{(s)}}$, where $\mathbf{L^{(s)}}$ is the $s$-th column of the Leontief matrix $\mathbf{L}$.

Let $A^s$ be the size of the avalanche (or cascade) triggered by a negative final-demand shock in sector $s$, and define $A^s$ to be equal to the number of sectors for which $\mathbf{\Delta x}<0$, i.e. sectors that are hit by the shock in $s$. By repeating this exercise for all sectors $s$, one can therefore characterize and study the avalanche size distribution $\{A^s, s=1,\dots,S\}$. Notice that our definition of avalanche and avalanche size is not affected by the size of the sector-specific shock.    

\subsubsection*{Model 2}

The input-output diffusion model described above assumes an exogenous shock on final demand and computes the impact on sectoral production keeping fixed, during the diffusion process, the magnitudes of inter-sectoral linkages and sectoral production. In the second diffusion model we study, we instead allow the magnitudes of economic input-output transactions to change during the propagation of the shock. 

Borrowing from Refs. \cite{Kinney2005,Lee2011}, suppose each node $s$ has a \textit{capacity} equal to its production $x_{s}$. Assume that both final demand and production are fixed during diffusion, a (negative) shock of size $f>0$ hits sector $s$, and this induces firms in the sector to modify their supplying and buying behaviors, thus leading to a change in inter-sectoral input-output linkages. This will affect all sectors that are linked to $s$ as buyers or sellers. If such an impact is too strong, also these sectors will be hit by the shock, thus resulting in a further change of their interlinkage magnitudes, and so on. When the reaction chain stops, i.e. no further sectors are hit by the shock, production is updated and we evaluate the size of the avalanche by counting the sectors eventually hit after the initial shock to sector $s$.

More precisely, suppose that, after the negative shock, output supply and input demand by sector $s$ is symmetrically decreased a fraction $0<f<1$. Consequently, the new link weights of sector $s$ become $z_{sj}^{*}=(1-f)z_{sj}$ and $z_{is}^{*}=(1-f)z_{is}$, where $j$ is any sector that uses commodity $s$ as input and $i$ is any sector from which $s$ buys inputs. In the next stage, every sector $h\neq s$, which is neighbor of $s$, evaluates the change in its total node strength: 
\begin{equation}
\Delta \sigma_h=\sum_k{(z_{hk}^{*}+z_{kh}^{*})}-\sum_k{(z_{hk}+z_{kh})} \label{eq:threshold}
\end{equation}
i.e. the change in the sum of all its incoming and outgoing link weights. If such a change exceeds a given share $0<c<1$ of its capacity $x_h$, then the sector is hit by the shock as well. It will therefore decrease its incoming and outgoing link weights by the proportion $f$ and transmit the shock farther away.     

Using the definitions of $z_{hk}^{*}$ and $z_{kh}^{*}$ in Eq. \eqref{eq:threshold}, one gets that the condition for a sector being hit by the initial shock becomes:
\begin{equation}
\frac{\sigma_h}{x_{s}} = \frac{\sigma_{h}^{in}}{x_{s}}+\frac{\sigma_{h}^{out}}{x_{s}} > \frac{c}{f} = \alpha,
\label{eq:threshold2}\end{equation}
where $\sigma_{h}^{in}$ and $\sigma_{h}^{out}$ are node in- and out-strength, i.e., respectively, the total value of the inputs bought by sector $h$ and the total value of sector $s$'s output used in the production processes of all other sectors.

Eq.~\eqref{eq:threshold2} implies, first, that a shock is transmitted to a neighboring node only when this sector is too exposed (relative to its capacity) to input-output relationships. Second, as already discussed by Ref. \cite{Lee2011}, shock propagation only depends on $\alpha$ and not on $c$ and $f$ separately. Nevertheless, the larger $\alpha$, the more likely a sector will absorb the shock. This happens the larger the resilience to shocks of nodes ($c$) and linkages ($f^{-1}$). Therefore, we interpret here $\alpha$ as a global measure of network resilience. Again, we are interested in the avalanche size distribution $\{A^s, s=1,\dots,S\}$, resulting from the diffusion dynamics starting from shocks occurring in any single sector.

\subsubsection*{Model 3}

The third and final diffusion model we study takes on board adaptation in production during shock propagation. Indeed, in the second model above, after a sector $s$ gets hit by a shock, the magnitude of its economic transactions decreases by $f$. This means that, as the diffusion process unfolds, the matrix $\mathbf{Z}$ keeps changing, but this does not have any effect on sector production, which keeps constant. In our third model, we account for dynamic production updating according to Eq.~\eqref{eq:IOM1}. This means that a sector that is hit by a shock has less to produce and therefore less to supply to other sectors. In this model, final demand remains fixed and everything works as before as far as shock propagation is concerned. 

As to production updating, assume that at some stage $\tau$ of the diffusion process, the system is characterized by the inter-sectoral weight matrix $\mathbf{Z}(\tau)$ and production vector $\mathbf{x}(\tau)$. At this point, assume that sector $h$ is hit by the shock. This results in the new weight matrix $\mathbf{Z}(\tau+1)$, which differs from $\mathbf{Z}(\tau)$ because its $h$-th row and column has been updated according to the rules $z_{hj}(\tau+1)=(1-f)z_{sh}(\tau)$ and $z_{ih}(\tau+1)=(1-f)z_{ih}(\tau)$. We then use Eq.~\eqref{eq:IOM3} and define the new production vector as:
\begin{equation}
\mathbf{x}(\tau+1)=(1-\mathbf{\Theta}(\tau+1))^{-1}\mathbf{d}=\mathbf{L(\tau+1)}\mathbf{d},
\label{eq:prod_updating}
\end{equation}
where $\mathbf{\Theta}(\tau+1)$ is the new technological coefficients matrix, whose generic entry reads $z_{ij}(\tau+1)/x_{ij}(\tau)$. This mechanism can be viewed as a self-fulfilling process where feedbacks arise and effects are reinforced. In this self-fulfilling process each update is incorporating previous updates. Updated production levels are then employed to evaluate if a shock hits a sector using Eqs.~\eqref{eq:threshold} and~\eqref{eq:threshold2}.

\subsection{Data\label{SubSec:Data}}

We calibrate the foregoing three models for EU countries using IO data tables provided by Eurostat. Data are available at: \url{http://epp.eurostat.ec.europa.eu/}, see the [SM] Section for more details. Tables give information on the economic transactions that sectors made by buying and supplying inputs in million Euros using 2-digit (division-level) NACE Rev. 1 classification. We employ year 2005 because this is the snapshot where the largest number of sectors can be observed ($S=59$) and, at the same time, can provide us with a picture of the pre-crisis conditions over which the propagation of shocks from financial to real sectors has been unfolding. Only four countries (Bulgaria, Cyprus, Latvia, and Malta) have been left out from the analysis due to data lacking. This leaves us with 22 countries \footnote{Austria, Belgium, Czech Republic, Denmark, Estonia, Finland, France, Germany, Greece, Hungary, Ireland, Italy, Lithuania, Luxembourg, Netherlands, Poland, Portugal, Romania, Slovenia, Spain, Sweden, UK.} for the analysis. 

We employ the data on inter-sectoral IO flows to build for each country $c=1,\dots,22$ an IO weighted-network matrix $\mathbf{Z}^c$. We also use data about intermediate and final demand to compute production and the Leontief inverse matrix as in Eq.~\ref{eq:IOM1}.

The topological properties of country-specific IO networks have been already studied in Refs. \cite{Slater1978,Blochl2011,McNerney2012}, from both a binary and weighted perspective. In the SM section, we report a short overview of the statistical features of the IO networks in our database.

\section{Results\label{Sec:Results}}

\subsection*{Model 1}
\begin{figure}[h!]
	\centering
	\includegraphics[width=0.45\textwidth]{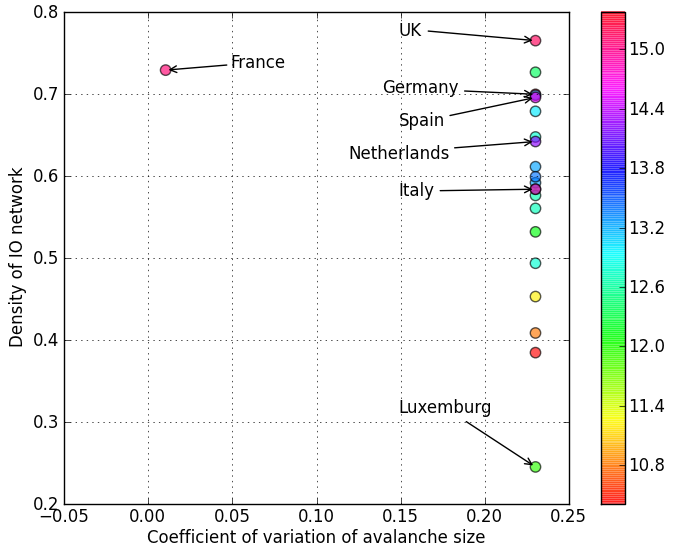}
	\caption{Model 1: Density of the input-output network (y-axis), coefficient of variation (ratio of sample standard deviation to sample mean) of avalanche size distribution (x-axis), logs of country GDP (ball color, see colormap), largest avalanche size (size   balls).}
	\label{fig:mod1_varcoeff}
\end{figure}

Independently of the size of the shock on final demand, diffusion in Model 1 triggers a very homogenous cascading process within most of EU countries. This can be seen in Fig. \ref{fig:mod1_varcoeff}, where for each country we plot the coefficient of variation (CoV) of country avalanche distribution $\{A^s, s=1,\dots,S\}$, defined as the ratio between standard deviation and mean, vs the density of the correspondent IO country network. To appreciate the extent of avalanches triggered by unit final-demand shocks, as well as the dependence on economic size, we draw each ball with a color proportional to the log of country GDP and a size proportional to the largest avalanche size, i.e. $\max_s\{A^s\}$. We can see that the CoV are all very small and homogeneous across all countries \footnote{This is because all countries (except France) experienced very large and homogeneous avalanche sizes, together with at least one avalanche of size 1. Conversely, all sectors in France triggered avalanches of size 54, with the exception of only an avalanche of size 57.}, indicating that all avalanche-size distributions are very concentrated, close to the maximum avalanche size possible, which is reached in all countries (see ball colors in the plot). 

Due to linearity of the diffusion process, coupled with fixed technological coefficients, we also notice that most of the avalanches were triggered by similar \textit{primary} sectors in all countries\footnote{For example, largest avalanches were often triggered by the Tobacco sector, whereas unit avalanche sizes were generated by uranium and thorium ores, and metal ores.}, independently on the size of the shock.

\begin{figure}[h!]
	\centering
	\includegraphics[width=0.45\textwidth]{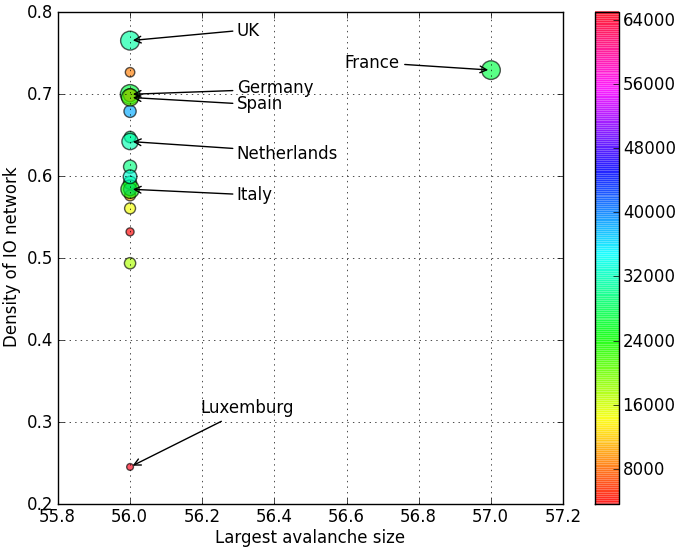}
	\caption{Model 1: Density of the input-output network (y-axis), largest avalanche size (x-axis), logs of country GDP (size of balls),  country GDP per capita (ball color, see colormap).}
	\label{fig:mod1_avsize}
\end{figure}

The fact that Model 1 was only able to generate homogeneous, very large avalanche sizes for all countries also implies the absence of any clear relationship between the size of the largest avalanche and country characteristics, such as density of its IO network, GDP per capita, or country size (see figure~\ref{fig:mod1_avsize}).

\subsection*{Model 2}
We now introduce the possibility that IO linkages get updated as propagation unfolds, after a sector capacity is hit by a negative shock. We study diffusion in IO networks using Model 2 in two extreme scenarios, i.e. high or small network resilience ($\alpha=c/f$). In the first scenario (high resilience), we set $f=0.6$ and $c=0.4$, whereas in the second one (low resilience) we set $f=0.7$ and $c=0.1$. Similar results hold also for other parameter-value combinations of $c$ and $f$. Note however that for values of $\alpha$ too large ($f\ll c$), that is for a shock too small and a capacity threshold too high, avalanches are of size zero throughout. So we focus on parameter constellations where $f>c$. 

\begin{figure}
	\centering
	\includegraphics[width=0.45\textwidth]{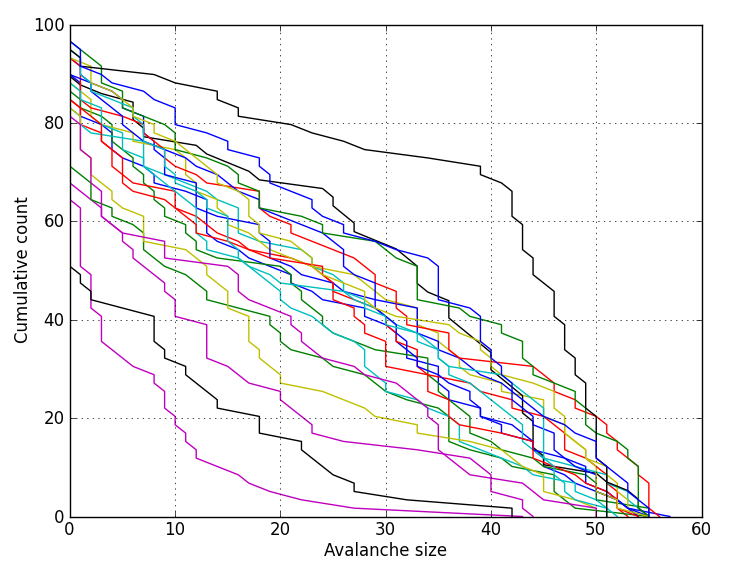}
	\caption{Model 2: Counter cumulative avalanche-size distributions for EU countries. High resilience ($f=0.6$ and $c=0.4$).}
	\label{fig:ASModel2I}
\end{figure}

\begin{figure}
	\centering
	\includegraphics[width=0.45\textwidth]{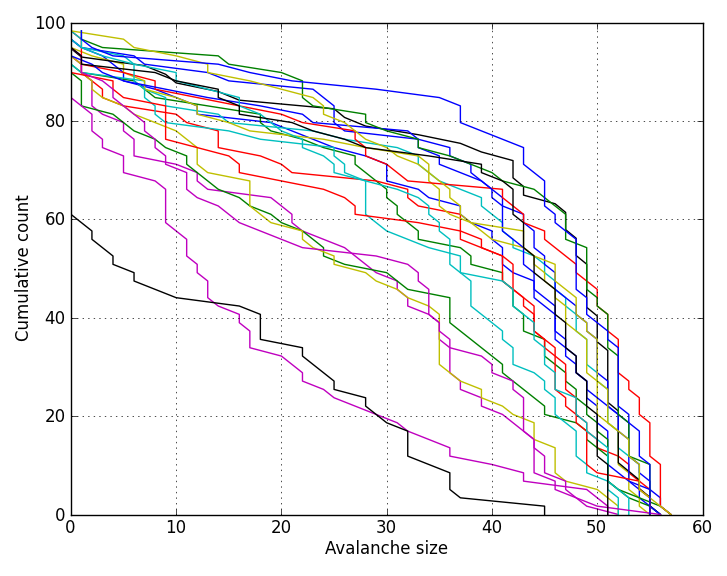}
	\caption{Model 2: Counter cumulative avalanche-size distributions for EU countries. Low resilience ($f=0.7$ and $c=0.1$).}
	\label{fig:ASModel2II}
\end{figure}

Our first result is that Model 2 generates very heterogeneous avalanche size distributions (both within and across countries), see Figures \ref{fig:ASModel2I} and \ref{fig:ASModel2II}. Note also that, as expected, a lower system resilience induces broader avalanche-size distributions, with more likely high-impact cascades.

\begin{figure}[h!]
	\centering
	\includegraphics[width=0.45\textwidth]{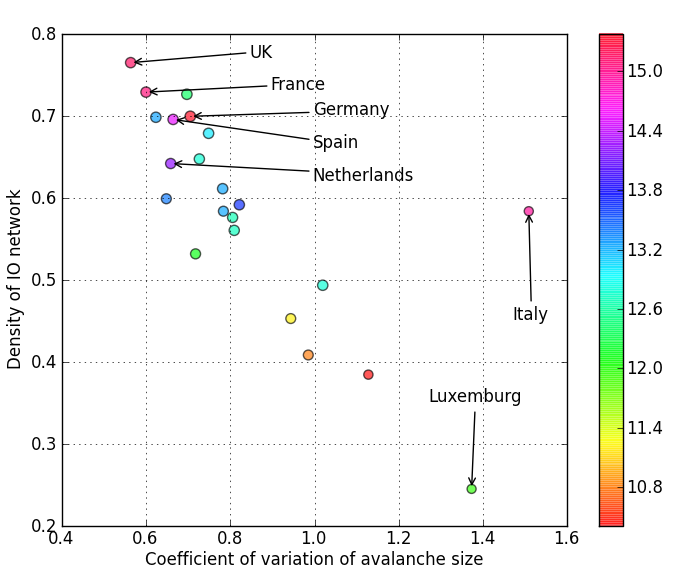}
	\caption{Model 2, High-Resilience Scenario: Density of the input-output network (y-axis), coefficient of variation (ratio of sample standard deviation to sample mean) of avalanche size distribution (x-axis), logs of country GDP (ball color, see colormap), largest avalanche size (size of balls).}
	\label{fig:mod2_varcoeff}
\end{figure}

\begin{figure}[h!]
	\centering
	\includegraphics[width=0.45\textwidth]{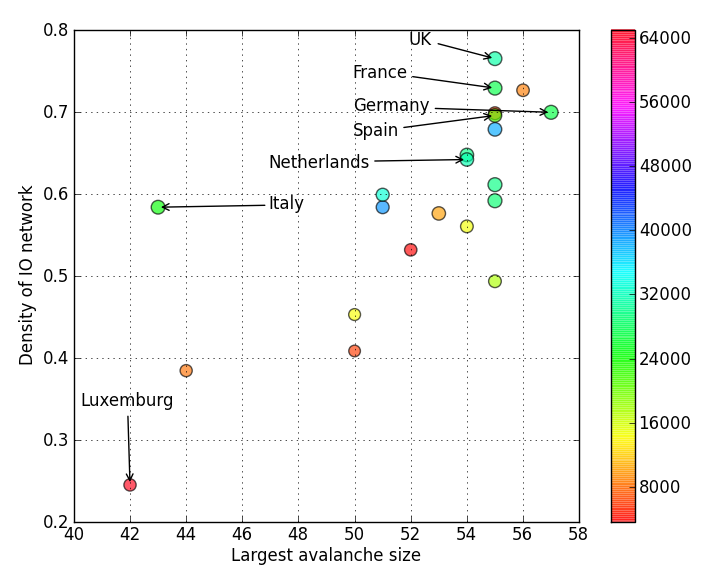}
	\caption{Model 2, High-Resilience Scenario: Density of the input-output network (y-axis), largest avalanche size (x-axis), logs of country GDP (size of balls),  country GDP per capita (ball color, see colormap).}
	\label{fig:mod2_avsize}
\end{figure}

The induced heterogeneity in avalanche-size distribution maps into interesting correlation patterns with country characteristics. As Figures \ref{fig:mod2_varcoeff} and \ref{fig:mod2_avsize} show for the high-resilience case (but similar results hold also for the low-resilience scenario), the more interconnected the IO network, the larger the biggest avalanche size and the smaller the CoV of the avalanche size distribution. Note also that the largest European countries in terms of their GDP typically experience the largest avalanches (with the exception of Italy), whereas there seem to be only a slightly negative relationship with country income.

This implies that countries with more interconnected input-output networks are more likely to experience stronger crises. These countries are likely to be large in terms of GDP, although not necessarily the richest ones. Therefore, what counts to induce larger avalanches is the development of the IO structure in terms of connectivity and not country income. However, small countries having a relatively less connected IO structure experience lower but more diverse avalanche sizes.

We turn now to investigate which sectors are more likely to trigger the largest avalanches (see [SM] for more details). In the high-resilience scenario, the sectors that triggered the largest avalanche sizes in most of the countries were wholesale (19 countries), other business services (19 countries), construction (18 countries), food and beverages (16 countries), and chemicals (14 countries). The ``financial-intermediation'' and ``insurance'' sectors triggered the largest avalanche sizes only in Luxembourg, although their impact was relevant throughout. Conversely, sectors that were more likely to trigger unit avalanche sizes were activities in the primary sector, except agriculture, such as forestry, fishing, coal and lignite peat, and metal ores.

In the low-resilience case, instead, the list of sectors capable of inducing the largest avalanche sizes considerably expands. The most common triggers of the largest avalanches are chemicals (21 countries), wholesale (19 countries), other business services (19 countries), construction (17 countries), electrical energy and gas (15 countries), hotels and restaurants (13 countries), and food and beverages (12 countries). As compared to the previous simulations, chemicals is now a trigger of the largest avalanches in seven more countries, electrical energy becomes a common trigger of the largest avalanches, and food and beverages becomes less common than before. In this setup, the countries that experienced the largest avalanches, covering 57 sectors, were France (triggered by chemicals, construction, and other business services), Germany (triggered by chemicals, other business services, and public administration), Greece (triggered by wholesale, and retail), Hungary (triggered by land transportation and food and beverages), and Spain (triggered by wholesale and chemicals). Other countries that experienced large avalanches of almost the totality of the economy (avalanches of size 56) were Belgium, Denmark, the Netherlands, and Slovenia.

\begin{figure}[h!]
	\centering
	\includegraphics[width=0.45\textwidth]{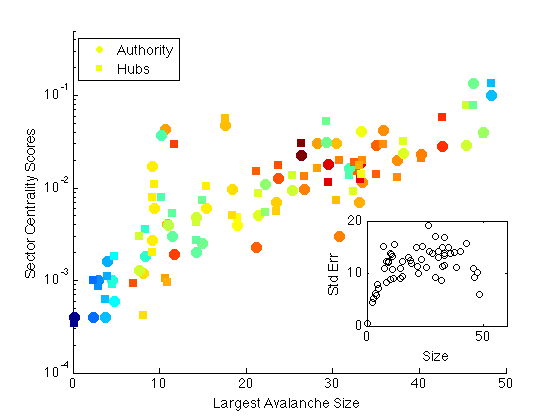}
	\caption{Model 2, High-Resilience Scenario: Cross-country averages of largest avalanche sizes vs. cross-country averages of sector centrality score. Markers colored proportionally to the cross-country standard error of largest avalanche size. Inset: cross-country standard error of largest avalanche size vs. average of largest avalanche size.}
	\label{fig:mod2_centrality}
\end{figure}

More generally, we find that sectors that are more globally central in the IO networks are also those triggering the largest avalanche sizes. To get a better feel for this result, Figure \ref{fig:mod2_centrality} plots, for the high-resilience scenario, cross-country averages of hubs and authority centrality scores (in log scale) \cite{Kleinberg1999} against cross-country averages of largest avalanche sizes. A strong positive correlation emerges. Note that a much weaker positive correlation emerges with \textit{local} sector centrality (as measured by sector in- and out-strength). This indicates that despite sectors get hit by a shock using local rules involving a sector in- and out-strength (i.e. their local centrality, see Eq. \ref{eq:threshold2}), the extent of the ensuing cascades mostly depends on the overall embeddedness of a sector in the IO network, which depends also on the centrality of all other sectors involved in a cascade.

Furthermore, in Figure \ref{fig:mod2_centrality} we color each observation proportionally to the cross-country standard errors associated with the average of largest avalanche sizes (on a blue to red range). It is easy to see that the smallest variations are obtained for small and big avalanche sizes, whereas the variability is higher for intermediate values of the avalanche size. This hints to an inverse U-shaped relation between cross-country averages and standard errors of largest avalanche sizes, which is confirmed by the inset of Figure \ref{fig:mod2_centrality}. Similar results hold also for the low-resilience scenario and suggest that whenever a sector is able to induce either a large or a small average largest avalanche size, then it also does so rather homogeneously across countries.      

\subsection*{Model 3}

\begin{figure}[h!]
	\centering
	\includegraphics[width=0.45\textwidth]{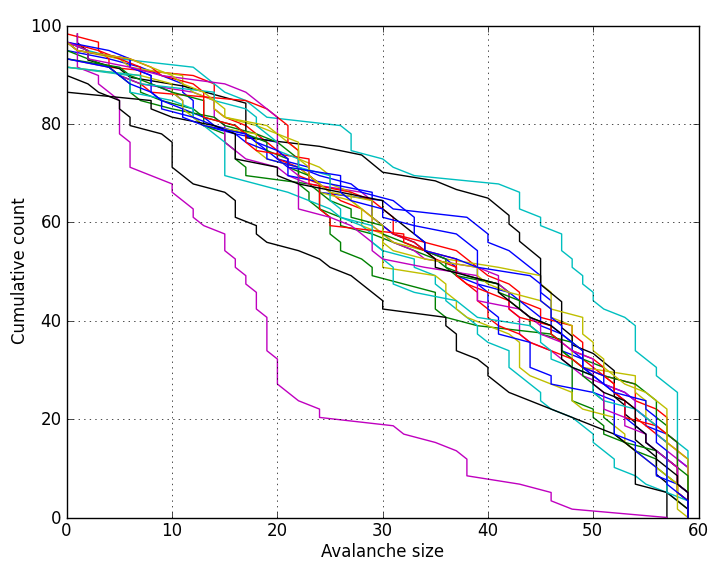}
	\caption{Model 3: Counter cumulative avalanche-size distributions for EU countries. High resilience ($f=0.6$ and $c=0.4$).}
	\label{fig:ASModel3I}
\end{figure}

\begin{figure}[h!]
	\centering
	\includegraphics[width=0.45\textwidth]{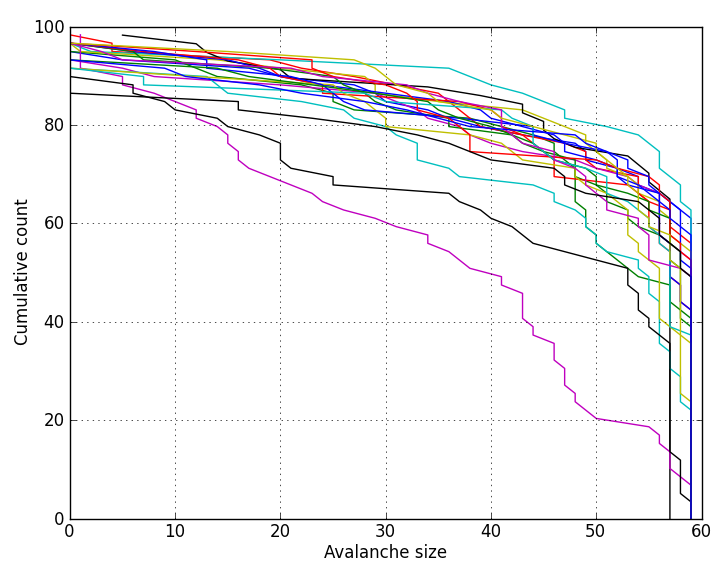}
	\caption{Model 3: Counter cumulative avalanche-size distributions for EU countries. Low resilience ($f=0.7$ and $c=0.1$).}
	\label{fig:ASModel3II}
\end{figure}

We now assume that, when hit by a shock, a sector adjusts not only the magnitudes of its connections within the IO network, but also the level of its production according to Eq. \eqref{eq:prod_updating}. Simulations show that this additional adaptation mechanism typically reinforces the strength and scope of the ensuing avalanches, making countries more vulnerable. At the same time, avalanche size distributions become more concentrated around large values, cf. Figures \ref{fig:ASModel3I} and \ref{fig:ASModel3II}. Therefore, Model 3 induces a cascading process which resembles that of Model 1, but with considerably more heterogeneity. The tendency toward more homogeneous and large cascades is due to the fact that, after adjusting production, sectors experience lower capacity thresholds, and thus shock propagation becomes easier. In other words, negative shocks resulting in production adjustments trigger a reinforcing mechanism wherein economies become weaker and more vulnerable, even if the shock is quite small, due to the coupled effect of linkage and production updating. 

In the high-resilience case, this implies that almost all countries experienced avalanches covering the entire economy. The only exception was Italy, experiencing avalanches of 57 sectors. All countries also experienced an increased number of avalanches of 58, 57 and 56 sectors.

As expected, the shift to the right of avalanche-size distributions, and their increased homogeneity, is more marked in the low-resilience case (Figure \ref{fig:ASModel3II}). A higher shock and a lower capacity translated into a higher frequency of avalanches covering the entire economy. Indeed, 50\% of countries were characterized by more than 30 sectors triggering avalanches for the entire economy. Among the common triggers of the largest avalanches we now also find agriculture and financial sectors. Also the number of avalanches of size larger than one increased, thus reducing the frequency of avalanches of size one.

\begin{figure}[h!]
	\centering
	\includegraphics[width=0.45\textwidth]{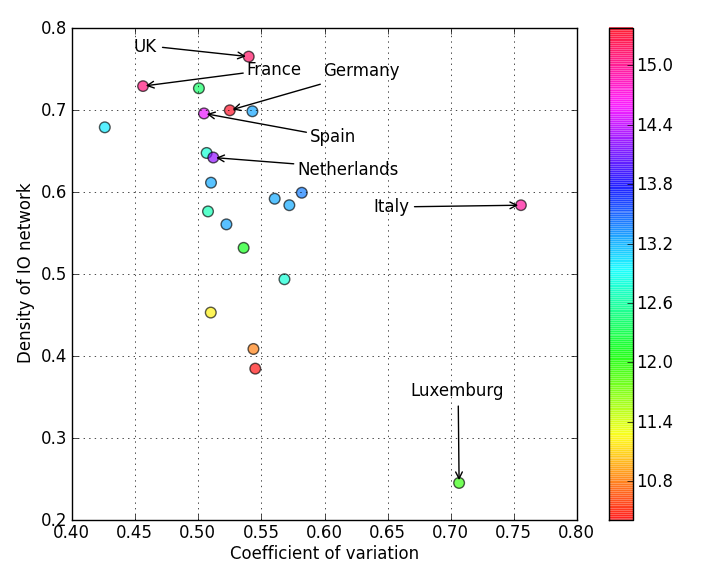}
	\caption{Model 3, High-Resilience Scenario: Density of the input-output network (y-axis), coefficient of variation (ratio of sample standard deviation to sample mean) of avalanche size distribution (x-axis), logs of country GDP (ball color, see colormap), largest avalanche size (size of balls).}
	\label{fig:mod3_varcoeff}
\end{figure}

\begin{figure}[h!]
	\centering
	\includegraphics[width=0.45\textwidth]{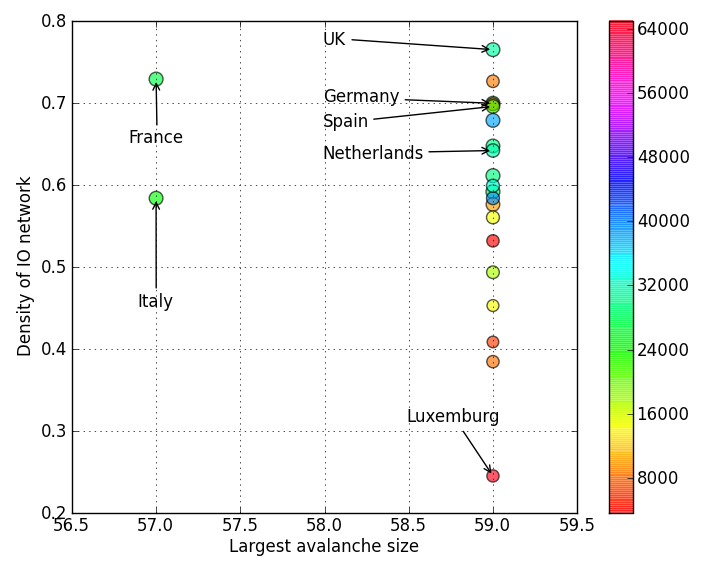}
	\caption{Model 3, High-Resilience Scenario: Density of the input-output network (y-axis), largest avalanche size (x-axis), logs of country GDP (size of balls),  country GDP per capita (ball color, see colormap).}
	\label{fig:mod3_avsize}
\end{figure}

Due to the fact that avalanche-size distributions are now very concentrated on their largest attainable values, the model with production updating does not feature robust correlation pattern between network density, country characteristics, and statistical properties of avalanche-size distributions, cf. Figures \ref{fig:mod3_varcoeff}-\ref{fig:mod3_avsize}. Note how the increase in the number of medium and large avalanches in all countries entails lower coefficients of variation. If any, a weak and negative relation is maintained between country size and coefficient of variation of avalanche size distribution.

\begin{figure}[h!]
	\centering
	\includegraphics[width=0.45\textwidth]{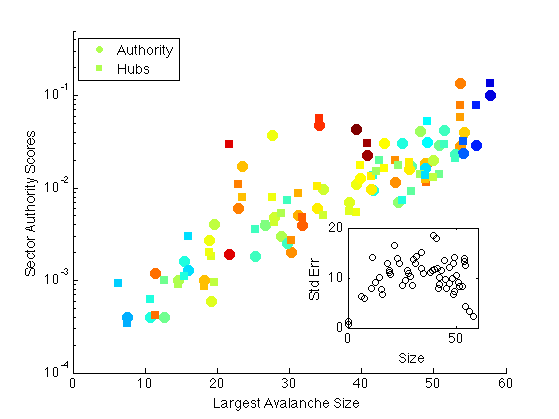}
	\caption{Model 3, High-Resilience Scenario: Cross-country averages of largest avalanche sizes vs. cross-country averages of sector centrality score. Markers colored proportionally to the cross-country standard error of largest avalanche size. Inset: cross-country standard error of largest avalanche size vs. average of largest avalanche size.}
	\label{fig:mod3_centrality}
\end{figure}

The marked shift to the right of avalanche-size distributions induced by production updating in all countries did not affect however the way in which different sectors trigger cascades in the economies under study. As discussed in details in the [SM] Section, a dominant role in generating the largest avalanches is still played by service and now the financial sectors. Furthermore, an even stronger positive relation between sector hubs and authority centrality in IO networks, and largest avalanche size, does emerge, see Figure \ref{fig:mod3_centrality}. This implies that, even when sectors update their production during propagation, their global centrality mostly explains their importance in channeling and amplifying the initial shock. Finally, as it happens in Model 2, the relation between cross-country average and standard errors of largest avalanche sizes still follows an inverse-U (see inset of Figure \ref{fig:mod3_centrality}).
 
\section{Conclusions\label{Sec:Conclusions}}
In this paper, we have investigated the mechanisms through which economic shocks are diffused and amplified by the input-output structure of national economies. We have studied three economically-meaningful diffusion models on networks, properly calibrated using real-world data for several European countries before the Great Depression. The models have been chosen so as to assess the role played by the nature of the shock and its impact on the economy. In particular, our models allow to evaluate the relative importance of a final demand- or production-driven shock, as well as the relevance of diffusion mechanisms involving, during propagation, the update of input-output technological interlinkages and/or sectoral production levels.

Simulation results show that, on the one hand, shocks affecting final demand without changing production and the technological relationships between sectors have on average a large but very homogeneous impact on the economy. On the other hand, when shocks change also the magnitudes of input-output across-sector interdependencies (and possibly sector production), the economy is subject to predominantly large but more heterogeneous avalanche sizes. In particular, heterogeneity is larger when sectors do not update production during propagation. 

Overall, the larger country size and the more globally central the sector in the networked input-output economy, the stronger its impact. This implies, from a policy perspective, that countries that are ``too big to fail'' are also more vulnerable to large economic shocks. Furthermore, our results suggest that the systemic importance of industrial sectors should not be evaluated only by looking at their economic size (e.g., in terms of value added or employees), but also at their position and embeddedness in the complex fabric of input-output relations.

The foregoing analysis can be extended in several directions. First, one might investigate the impact of shocks not only in terms of avalanche size but also in terms of avalanche intensity. Indeed, in the exercises above we have focused our attention on avalanche-size distributions in general, and on the largest avalanches in particular. This has been done because we were interested in assessing the very possibility that a small shock can propagate or not in the entire economy. More generally, one might also want to target the total change in sectoral production induced by the shocks. Second, one can play with alternative models of diffusion on networks, possibly involving some (more sophisticated) micro-foundation in terms of firm production behavior, in line with previous work \cite{Bak_etal_1993,Acemoglu2012,Acemoglu2013}.       

\newpage \cleardoublepage

\newpage \clearpage


\onecolumngrid
\appendix*

\renewcommand\thefigure{\Roman{figure}} 
\renewcommand{\theequation}{\Roman{equation}}

\setcounter{figure}{0} 
\setcounter{equation}{0} 

\begin{center}\textbf{SUPPLEMENTAL MATERIAL}\end{center}
\bigskip

\section*{Data}
\noindent We study input-output (IO) structure in EU countries using symmetric input-output data tables provided by Eurostat. Data are available at: \url{http://epp.eurostat.ec.europa.eu/}. Tables give information on the economic transactions that sectors made by buying and supplying inputs in million Euros using 2-digit (division-level) NACE Rev. 1 classification. In this classification the main criteria applied in delineating sectors and divisions concern the following characteristics of the activities of production units: the character of the goods and services produced; the uses to which the goods and services are put; and the inputs, the process and the technology of production. One NACE code is assigned to each unit recorded in statistical business registers, according to its principal economic activity. The principal activity is the activity which contributes most to the value added of the unit. See NACE Rev. 2. Statistical classification of economic activities in the European Community. Eurostat, Methodologies and Working papers ASSN 1977-0375, p. 27.

This classification uses the a top-down method, which follows a hierarchical principle: the classification of a unit at the lowest level of the classification must be consistent with the classification of the unit at the higher levels of the structure. To satisfy this condition the process starts with the identification of the relevant highest level and progresses down through the levels of the classification (see NACE Rev. 2. Statistical classification of economic activities in the European Community. Eurostat, Methodologies and Working papers ASSN 1977-0375 (http://epp.eurostat.ec.europa.eu/cache/ITY\_OFFPUB/KS-RA-07-015/EN/KS-RA-07-015-EN.PDF)).

We employ year 2005 because this is the snapshot where the largest number of sectors can be observed ($S=59$) and, at the same time, can provide us with a picture of the pre-crisis conditions over which the propagation of shocks from financial to real sectors has been unfolding. Only four countries (Bulgaria, Cyprus, Latvia, and Malta) have been left out from the analysis due to data lacking. This leaves us with 22 countries: Austria, Belgium, Czech Republic, Denmark, Estonia, Finland, France, Germany, Greece, Hungary, Ireland, Italy, Lithuania, Luxembourg, Netherlands, Poland, Portugal, Romania, Slovenia, Spain, Sweden, UK. 

\section*{Input-Output Model}

\noindent Input-output analysis provides the tools to analyze the inter-sectoral dependencies and the impact of a sector on the economy. The tools that researchers have used in this approach have been linkage measures, multipliers, and structural decomposition. These tools rely on the input-output relationships framed in the model proposed by \citet{Leontief1936} where sectoral output is a function of the input requirements from all sectors in the economy and final demand of an exogenous sector constituted by households, government, exports, and capital investment. Backward linkage and output multipliers measure the output required to satisfy additional final demand. A sector with a high output multiplier is one that has a high impact in terms of the total dollar value of output generated throughout the economy. Structural decomposition has studied the nature of changes in the inter-sectoral structure according to changes in technology and in final demand \citep{Dietzenbacher1998}. Despite the fact that input-output analysis has studied the connections and interdependencies between sectors, it has not explored the emergence and nature of cascading effects or avalanches from one sector to another explicitly. Our investigation fills this gap by complementing the analysis with network diffusion models.

We use the intermediate demand table to compute the technical coefficients matrix following Leontief's \cite{Leontief1936} input-output model, and to construct the input-output network. In the input-output model, total output of a sector is expressed as a function of the demand for the different commodities produced in the economy. Production is defined as:

\begin{equation}\label{eq:IOM1}
\mathbf{x = Z1+d}
\end{equation}
where $\mathbf{x}$  is the $S \times 1$ column vector of output, $\mathbf{Z}$ is the inter-industry flows $S \times S$ matrix or input-output matrix, in which $z_{ij}$ represent inter-industry sales by sector i to sector j, $\mathbf{1}$ is a column vector of ones, and $\mathbf{d}$ is the $S \times 1$ column vector of final demand.

Define the technical coefficients as the ratio of input supplied by $i$ and bought by $j$ over the output of sector $j$, $\theta_{ij}=z_{ij}/x_{j}$, and substitute the $\mathbf{Z}$ matrix in equation~\ref{eq:IOM1} for $\theta_{ij}$ to obtain:

\begin{equation}\label{eq:IOM2}
\mathbf{x = \Theta x+d}
\end{equation}
where $\mathbf{\Theta}=[\theta_{ij}]=z_{ij}/x_{j}$ is the $S \times S$ matrix of technical coefficients.

Solving for $\textbf{x}$ yields:

\begin{equation}\label{eq:IOM3}
\mathbf{x = (I - \Theta)^{-1} d = Ld}
\end{equation}
where $\mathbf{L}=\mathbf{(I-\Theta)}^{-1} = [l_{ij}]$ is an $S \times S$ matrix known as the Leontief inverse or the total requirements matrix. Technological coefficients are input per output, thus give a measure of the requirements of a sector to produce, see Ref. \cite[][p. 275-278]{Leontief1949}.

We compute total production of all individual sectors as a function of final demand, once we know the magnitudes of the technical coefficients following equation~\eqref{eq:IOM3}. Input-output tables from Eurostat database give the intermediate demands and final demand, which covers the information required to compute production and the Leontief inverse. Given this information we can evaluate the impact of sector $i$ on aggregate production and the production of each sector. This impact is defined as the change in production of sectors needed to compensate a change in final demand of sector $i$. This is computed applying equation~\eqref{eq:IOM4} defined below. Thus, a shock on final demand could be viewed as a change in government expenditure, or a change in the international trade performance. The change in production due to a change of final demand from $\mathbf{d^{0}}$ to $\mathbf{d^{1}}$ is given by:

\begin{equation}\label{eq:IOM4}
\mathbf{\Delta x = x^{1}-x^{0}} = \mathbf{L(d^{1} - d^{0})} = \mathbf{L\Delta d.} 
\end{equation}

A shock on final demand (an additional unit of final demand) of a sector translates into a change in production of the other sectors of the economy in different magnitudes depending on the input-output relationships. These input-output relationships shape the linkages through which the effect is spread, changing the flows of resources both to supply and buy inputs. 

\section*{Input-Output Networks: Topological Properties}

We now study the topological structure of input-output networks as defined in the Data and Methods section of the paper. Each input-output network refers to a country in 2005, and features 59 nodes, corresponding to the sector classification. One exception is the French input-output network, which had 57 non-isolated nodes, because there are two sectors disconnected from the rest of the economy: (i) uranium and thorium ores; (ii) recovered secondary raw materials. We refer the reader to Refs. \cite{Slater1978,Blochl2011,McNerney2012} for a more in-depth analysis. Here we just want to report on some interesting properties that can help us in better understanding the results of diffusion exercises we perform in the paper.

Table~\ref{table:Density} features some simple descriptive statistics about country IO networks: (i) network 
density (i.e. the ratio between the actual number of links and the number of links in place if all nodes were connected among them); (ii) bilateral density, defined as the ratio of reciprocated links over total number of links; (iii) network diameter (i.e., the largest shortest path between any two nodes); and (iv) average path length (i.e. average number of steps along the shortest paths for all possible pairs of nodes in the network). 

We observe that in all countries bilateral density is lower than overall network density, which is expected as production chains have a certain sequential order and not all sectors require each others' inputs. Notice that density of IO networks can be related to the level of development of the economic system. The idea is that the more developed an economy is, the more connections there will be, therefore the density will be higher. All networks display low diameters and average path length as expected. This has important implications for the spread of information or resources across the network, as the spread will be faster on networks with low average path length \citep{Newman2003}. Results showed that for every network the average path length is around one. This indicates that the networks are very responsive in terms of the spread of resources.

\singlespacing

\begin{table}[htbp]
\caption{EU Country Input-Output Networks: Descriptive Statistics}
\label{table:Density}
\begin{center}
	\begin{tabular}[c]{ccccc}
	\hline \hline
	\textbf{} &  & \textbf{Bilateral} & & \textbf{Average}\\ 
	\textbf{Countries} & \textbf{Density} & \textbf{Density} & \textbf{Diameter} & \textbf{Path Length}\\ \hline 
	Austria & 0.61 & 0.45 & 2 & 1.17\\ 
	Belgium & 0.59 & 0.43 & 2 & 1.19\\ 
	Czech Republic & 0.57 & 0.40 & 2 & 1.25\\ 
	Denmark & 0.67 & 0.55 & 3 & 1.16\\ 
	Estonia & 0.38 & 0.22 & 3 & 1.40\\ 
	Finland & 0.64 & 0.52 & 3 & 1.20\\ 
	France & 0.72 & 0.63 & 3 & 1.14\\ 
	Germany & 0.69 & 0.53 & 2 & 1.10\\ 
	Greece & 0.49 & 0.29 & 2 & 1.28\\ 
	Hungary & 0.72 & 0.65 & 2 & 1.17\\ 
	Ireland & 0.58 & 0.45 & 2 & 1.17\\ 
	Italy & 0.58 & 0.45 & 2 & 1.17\\ 
	Lithuania & 0.40 & 0.24 & 2 & 1.33\\ 
	Luxembourg & 0.24 & 0.10 & 3 & 1.57\\ 
	Netherlands & 0.64 & 0.47 & 3 & 1.15\\ 
	Poland & 0.69 & 0.58 & 2 & 1.09\\ 
	Portugal & 0.55 & 0.37 & 2 & 1.22\\ 
	Romania & 0.53 & 0.38 & 3 & 1.21\\ 
	Slovenia & 0.45 & 0.28 & 3 & 1.36\\ 
	Spain & 0.69 & 0.53 & 2 & 1.10\\ 
	Sweden & 0.59 & 0.50 & 2 & 1.10\\ 
	UK & 0.76 & 0.67 & 2 & 1.05\\
	\hline \hline
	\end{tabular}
\end{center}
\end{table}

\begin{figure}[h!]
	\centering
	\includegraphics[scale=0.5]{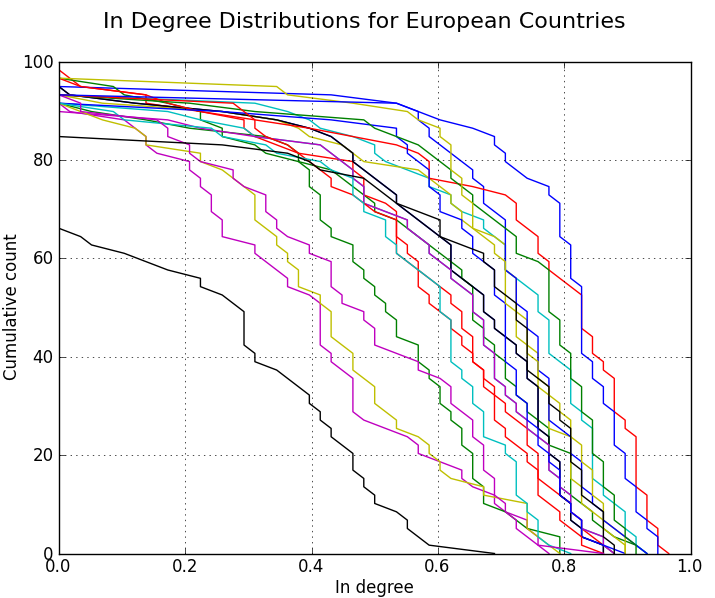}
	\caption{Counter cumulative distribution functions (1-cdf) for indegree values for each country}
	\label{fig:Figure1}
\end{figure}
\begin{figure}[h!]
	\centering
	\includegraphics[scale=0.5]{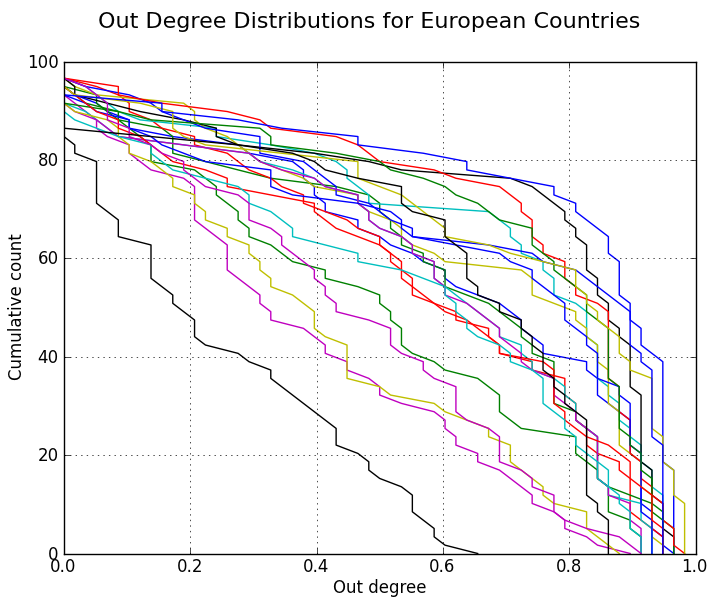}
	\caption{Counter cumulative distribution functions (1-cdf) for outdegree values for each country}
	\label{fig:Figure2}
\end{figure}

We now move to studying (in and out) degree distributions. In IO networks, node in-degree measures the number of sectors from which a sector buys inputs to be used in its production process, whereas out-degree counts the number of sectors that buy inputs from the node under consideration. We normalize degrees by the maximum possible degree $S-1$ to get a measure of degree centralization ranging in the unit interval. We find that degree distributions for European countries are highly negatively skewed, where the majority of the degrees were high in most of the countries (see Figures \ref{fig:Figure1} and \ref{fig:Figure2}). Exceptions are Estonia, Lithuania and Luxembourg which had higher frequency of small and medium in-degree values and Estonia, Greece, Lithuania, Luxembourg, Romania, and Slovenia which had more medium and small out-degrees, or even more homogeneous distributions. Luxembourg stands out from the rest of the countries by displaying a predominance of very small values.

\begin{figure}[h!]
	\centering
	\includegraphics[scale=0.5]{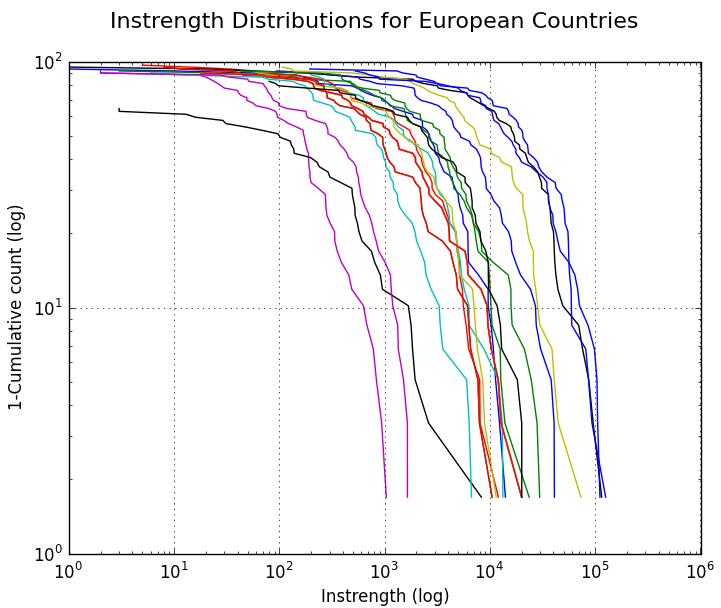}
	\caption{Counter cumulative distribution of node in-strength for each country (log-log plot).}
	\label{fig:Instrength}
\end{figure}

\begin{figure}[h!]
	\centering
	\includegraphics[scale=0.5]{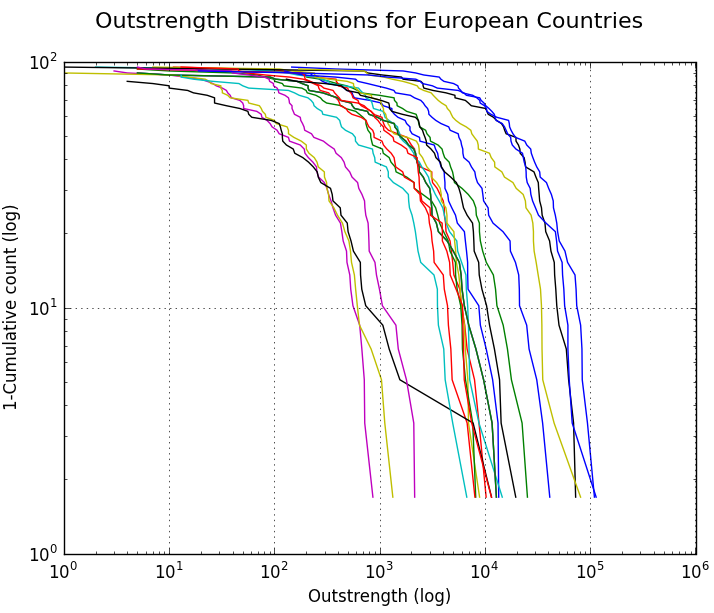}
	\caption{Counter cumulative distribution of node out-strength for each country (log-log plot).}
	\label{fig:Outstrength}
\end{figure}

We also studied cross-country distributions of in- and out-strength, defined respectively as the row and column sum of the weighted adjacency matrix of technological coefficients $\mathbf{Z}^c$. As opposed to degrees, strength distributions for most of the European countries are highly but \textit{positively} skewed, with quasi-Pareto steep upper tails. Some countries had skewer distributions than others, but, in general, low strengths predominate (see Figures \ref{fig:Instrength} and \ref{fig:Outstrength}).

These results, together with the degree distributions, point to the fact that the sectors in each country have many connections but that most of these connections are very weak.

Finally, we asked whether IO country networks are disassortative, as for example it happens in the International Trade Network \cite{Fagiolo2009pre}. We have computed binary and weighted network assortativity in two ways. First, following Ref. \cite{Newman2002}, we have calculated link-wise degree and strength assortativity, by correlating degrees or strengths of the two nodes lying at both sides of a link. Additionally, we have calculated the linear correlation coefficient between degree and node average nearest-neighbor degree (ANND); as well as correlation between node strength and node average nearest-neighbor strength (ANNS) in each network (see Ref. \cite{Barrat2004pnas,Barthelemy2005} for more details).

\begin{figure}[h!]
	\centering
	\includegraphics[scale=0.5]{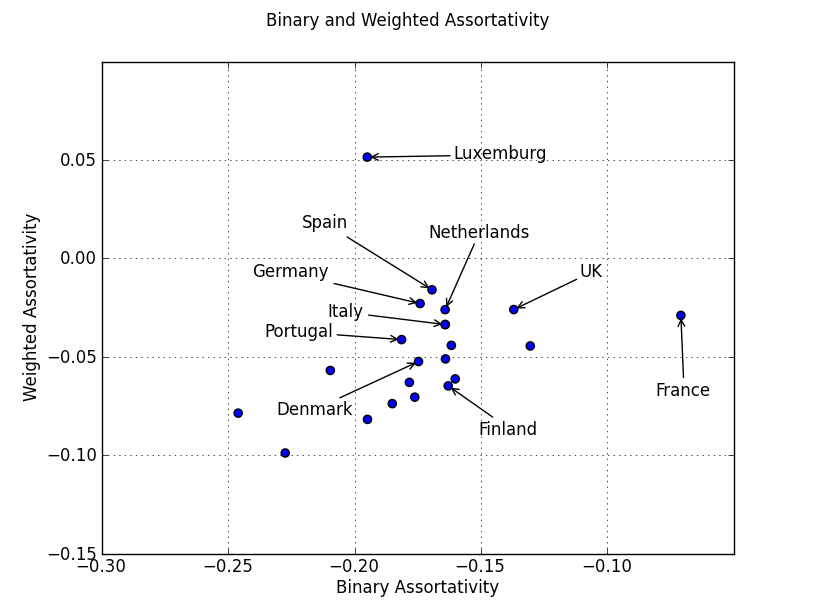}
	\caption{Binary Assortativity vs Weighted Assortativity. Binary Assortativity: Link-wise correlation between node degrees. Weighted Assortativity: Link-wise correlation between node strengths.}
	\label{fig:Assortativity}
\end{figure}

Results provide contrasting evidence. If assortativity is measured in terms of link-wise correlation between degrees or strengths, we can observe that the input-output network in every country is disassortative, although the coefficients remain small (see figure \ref{fig:Assortativity}). On the contrary, if we measure assortativity with the correlation between degrees/strengths and ANND/ANNS, all country networks exhibit some assortativity. This is in sharp contrast with similar literature on e.g. international trade \cite{Fa08} and can be explained by the peculiar hierarchical nature of IO networks \cite{McNerney2012}.

\begin{figure}[h!]
	\centering
	\includegraphics[scale=0.5]{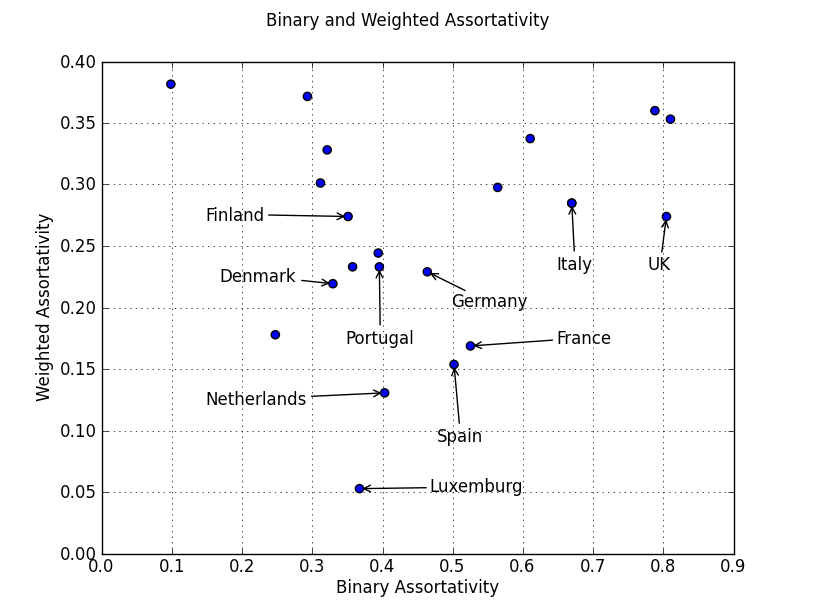}
	\caption{Binary Assortativity vs Weighted Assortativity. Binary Assortativity: Linear correlation coefficient between degree and node average nearest-neighbor degree (ANND). Weighted Assortativity: correlation between node strength and node average nearest-neighbor strength (ANNS).}	\label{fig:AssortativityI}
\end{figure}


\section*{Model 2: Inter-Sectoral Analysis}

This section discusses in more detail the relationship between sectoral structure and avalanche size distributions in different countries when Model 2 is used to simulate shock diffusion. Results on the sector that triggered the largest avalanches in most of the countries pointed out that the services provided by the units in sector other business services are essential for the functioning of the enterprises that use them in most of the economic activities, directly and indirectly, along the input-output network. When this sector is hit by a shock, its capacity to spread it to other sectors is high. This is because other business services has a central role in the European economies. This fact is reflected on high, if not the highest, global centrality measured as the authority and hub scores of the sector \cite{Kleinberg1999}. 

Other business services is a good authority and a good hub in most of the input-output networks meaning that it uses and, at the same time, supplies essential services for enterprises of many economic activities. These services include market research, opinion polls, computer services and services provided by means of office machines, employment agencies and security services, and debt collecting agencies, among others. Similarly, in Luxembourg the sector financial intermediaries and closely related services such as services auxiliary to financial intermediation that triggered the largest avalanches played a central role in the economy by providing essential services to most to the units in the other economic activities, and using the output of most of the other sectors as inputs to provide such services. This is too reflected on high global centralities (see table~\ref{table:AH}). For Greece, Hungary, Lithuania, Poland, Spain, and Sweden, the sector other business services was within the ten best authorities and hubs, although it did not have the highest score.

\begin{table}[htbp]
\caption{Authorities and Hubs}
\label{table:AH}
\begin{center}
\scalebox{0.75}{
	\begin{tabular}{|c|p{8cm}|p{8cm}|}
	\hline
	\textbf{Countries} & \textbf{Authorities} & \textbf{Hubs}\\ \hline
	Austria & Other business services and electrical energy & Other business services and electrical energy\\ \hline
	Belgium & Other business services and chemicals & Other business services and wholesale\\ \hline
	Czech Republic & Construction and other business services & Construction and other business services\\ \hline
	Denmark & Supporting and auxiliary transport services and other business services & Water transport and wholesale\\ \hline
	Estonia & Radio, television and communication and other business services & Radio, television and communication and electrical machinery\\ \hline
	Finland & Radio, television and communication and other business services & Radio, television and communication and construction\\ \hline
	France & Other business services and services auxiliary to financial intermediation & Other business services and wholesale\\ \hline
	Germany & Motor vehicles trailers and semi-trailers and other business services & Motor vehicles trailers and semi-trailers and other business services\\ \hline
	Greece & Agriculture and coal and lignite peat & Food and beverages and construction\\ \hline
	Hungary & Radio, television and communication and electrical machinery & Electrical machinery and motor vehicles trailers and semi-trailers\\ \hline
	Ireland & Other business services and chemicals & Other business services and chemicals\\ \hline
	Italy & Other business services and chemicals & Other business services and chemicals\\ \hline
	Lithuania & Crude petroleum and natural gas and electrical energy & Chemicals and electrical energy\\ \hline
	Luxembourg & Services auxiliary to financial intermediation and financial intermediation & Services auxiliary to financial intermediation and financial intermediation\\ \hline
	Netherlands & Other business services and construction & Other business services and construction\\ \hline
	Poland & Agriculture and food and beverages & Agriculture and food and beverages\\ \hline
	Portugal & Construction and other non-metallic mineral products & Construction and other business services\\ \hline
	Romania & Agriculture and food and beverages & Agriculture and food and beverages\\ \hline
	Slovenia & Construction and other non-metallic mineral products & Construction and other business services\\ \hline
	Spain & Construction and other non-metallic mineral products & Construction and  real-estate\\ \hline
	Sweden & $R\&D$ and real-estate & $R\&D$ and trade maintenance and repair services of motor vehicles\\ \hline
	UK & Other business services and construction & Other business services and construction \\
	\hline
	\end{tabular}
	}
\end{center}
\end{table}

\begin{table}[h!]
\caption{Model 2, High Resilience: Sectors that triggered the largest and smallest avalanches in each country.}
\label{table:LAI2}
\begin{center}
\scalebox{0.75}{
	\begin{tabular}[c]{|c|p{9cm}|c|p{8cm}|c|}
	\hline
	 & \multicolumn{2}{|c|}{\textbf{Largest Avalanche}} & \multicolumn{2}{|c|}{\textbf{Smallest Avalanche}}\\ \hline
	\textbf{Countries} & Triggering sector & Size & Triggering sector & Size\\ \hline
	Austria & Wholesale, construction, electrical energy, food and beverages, other business services & 55,53,52,50 & Fishing, coal, tobacco, wearing apparel & 0 \\ \hline
	Belgium & Chemicals, wholesale, food and beverages, other business services & 55,50 & Activities in the primary sector except agriculture & 0\\ \hline
	Czech Republic & Land transport, food and beverages, wholesale, other business services, chemicals & 53-51 & Activities in the primary sector, $R\&D$ & 0\\ \hline
	Denmark & Food and beverages, agriculture, construction, wholesale, chemicals, other business services & 55-51 & Activities in the primary sector except agriculture & 0\\ \hline
	Estonia & Wholesale, construction, other business services, food and beverages, land transport & 44-40 & Activities in the primary sector except agriculture & 0\\ \hline
	Finland & Food and beverages, paper, wholesale, construction, other business services & 54-52 & Crude petroleum, uranium and thorium ores, tobacco, wearing apparel & 0\\ \hline
	France & Other business services, chemicals, construction, food and beverages, wholesale & 55,54,53 & Forestry, fishing, uranium, metal ores, tobacco & 0 \\ \hline
	Germany & Chemicals, other business services, machinery and equipment, motor vehicles, construction, wholesale & 57,53,52 & Fishing, metal ores, tobacco, leather & 0\\ \hline
	Greece & Wholesale, food and beverages, construction, retail, other business services & 55,48,47 & Forestry, fishing, crude petroleum, metal ores, tobacco, and manufactures & \\ \hline
	Hungary & Food and beverages, coke and refined petroleum, electrical energy, wholesale, other business services & 56,55 & Fishing, metal ores, tobacco, water transport & 0 \\ \hline
	Ireland & Construction, other business services, chemicals, wholesale, hotel and restaurants & 51,50,48 & Fishing, wearing apparel, leather  & 0 \\ \hline
	Italy & Other business services, construction, chemicals, food and beverages, wholesale & 43,27,22,19 & Activities in the primary sector except agriculture and manufactures & 0 \\ \hline
	Lithuania & Chemicals, wholesale, electrical energy, food and beverages, retail & 50,48,45 & Forestry, Fishing, coal and lignite peat, leather & 0 \\ \hline
	Luxembourg & Construction, other business services, financial, wholesale, retail & 42,32,27 & Activities in the primary sector except agriculture and manufactures & 0 \\ \hline
	Netherlands & Chemicals, construction, coke and refines petroleum, electrical energy, other business services & 54,53,52,51 & Forestry, fishing, coal and lignite peat, metal ores, tobacco & 0 \\ \hline
	Poland & Electrical energy, chemicals, wholesale, retail, land transport & 55,54 & Tobacco, membership organisation services, fishing, leather  & 0,1\\ \hline
	Portugal & Wholesale, other business services, public administration, chemicals & 54,52 & Activities in the primary sector except agriculture & 0 \\ \hline
	Romania & Electrical energy, food and beverages & 52,51 & Recovered secondary raw materials, trade maintenance and repair services of motor vehicles, air transport & 0 \\ \hline
	Slovenia & Wholesale, construction & 50 & Activities in the primary sector except agriculture & 0 \\ \hline
	Spain & Chemicals, wholesale, hotels and restaurants, other business services & 55,52,50 & Forestry, fishing, metal ores, tobacco & 0 \\ \hline
	Sweden & Trade maintenance and repair services of motor vehicles, wholesale, retail, other business services & 51 & Activities in the primary sector except agriculture & 0 \\ \hline
	UK & Wholesale, other business services, construction, food and beverages, real-estate & 55,54 & Tobacco, leather & 0 \\
	\hline
	\end{tabular}
}
\end{center}
\end{table}

\begin{table}[h!]
\caption{Model 2, Low Resilience: Sectors that triggered the largest and smallest avalanches in each country.}
\label{table:LAII2}
\begin{center}
\scalebox{0.75}{
	\begin{tabular}[c]{|c|p{9cm}|c|p{8cm}|c|}
	\hline
	 & \multicolumn{2}{|c|}{\textbf{Largest Avalanche}} & \multicolumn{2}{|c|}{\textbf{Smallest Avalanche}}\\ \hline
	\textbf{Countries} & Triggering sector & Size & Triggering sector & Size\\ \hline
	Austria & Wholesale, chemicals, electrical energy, hotel and restaurants & 55-54 & Coal, tobacco, Fishing & 0-1\\ \hline
	Belgium & Wholesale, chemicals, supporting and auxiliary transport services, construction  & 56-55 & Forestry, fishing, coal, crude petroleum, uranium and thorium ores & 0-1\\ \hline
	Czech Republic & Chemicals, electrical energy, wholesale, land transport, other business services & 55-54 & Crude petroleum, uranium and thorium ores, metal ores, other mining and quarrying products, fishing & 0-1\\ \hline
	Denmark & Food products and beverages, chemicals, construction, wholesale, other business services & 56-55 & Coal, crude petroleum, uranium and thorium ores, metal ores, leather & 0-1\\ \hline
	Estonia & Wholesale,real estate, other business services, construction  & 52,49,48,46 & Metal ores, other mining and quarrying products, recovered secondary raw materials, $R\&D$& 0\\ \hline
	Finland & Food and beverages, electrical energy, real estate, chemicals, other business services  & 55,54,52 & Crude petroleum, uranium and thorium ores, tobacco products, fishing & 0-1\\ \hline
	France & Chemicals, construction, other business services, coke refined petroleum, electrical energy gas  & 56,55,54,53 & Metal ores, tobacco, coal and lignite peat  & 0,1 \\ \hline
	Germany & chemicals, other business services, public administration, electrical energy, wholesale, basic metals & 57,56,55 & Metal ores, tobacco, fishing & 0-1 \\ \hline
	Greece & Wholesale, retail trade, chemicals, real-estate, electrical energy  & 57,56,54,52 & Fishing, metal ores, tobacco, office machinery & 0 \\ \hline
	Hungary & Land transport, food and beverages, coke and refined petroleum, chemicals, other non-metallic mineral products, electrical energy, construction, wholesale, other business services & 57,56,55 & Metal ores, fishing, coal and lignite peat, recovered secondary raw materials & 0,1 \\ \hline
	Ireland & Construction, wholesale, hotel and restaurants, chemicals, other business services & 53,52 & Leather, tobacco, recovered secondary raw materials & 0,1 \\ \hline
	Italy & Construction, other business services, chemicals, wholesale, food and beverages & 51,50,49 & Fishing, tobacco, textiles, wearing apparel, leather & 0 \\ \hline
	Lithuania & Chemicals, wholesale, retail, electrical energy, construction, food and beverages & 52,51,50,47 & Coal and lignite peat, office machinery, air transport  & 0 \\ \hline
	Luxembourg & Construction, other business services, chemicals, financial, wholesale & 45,37,36 & Activities in the primary sector except agriculture, and basic manufactures & 0 \\ \hline
	Netherlands & Electrical energy, coke and refined petroleum, chemicals, construction, agriculture, other business services & 56,55,53 & Forestry, fishing, coal and lignite peat, metal ores, tobacco & 0,2 \\ \hline
	Poland & Chemicals, electrical energy, wholesale, retail, land transport  & 55,53 & Fishing, tobacco, leather & 1,3 \\ \hline
	Portugal & Chemicals, wholesale, hotels and restaurants, other business services  & 55,54 & Activities in the primary sector except agriculture & 0 \\ \hline
	Romania & Food and beverages, electrical energy, hotels and restaurants, chemicals  & 51,52 & Recovered secondary raw materials, trade maintenance and repair services of motor vehicles, fishing, air transport & 0,1 \\ \hline
	Slovenia & Wholesale, construction, food and beverages, chemicals & 56,50,48 & Activities in the primary sector except agriculture & 0 \\ \hline
	Spain & Chemicals, wholesale, hotels and restaurants, construction, other business services & 57,56,55 & Tobacco & 0 \\ \hline
	Sweden & Chemicals, construction, trade maintenance and repair services of motor vehicles, wholesale, retail, other business services & 51,50 & Fishing, crude petroleum and natural gas, uranium and thorium ores & 0 \\ \hline
	UK & Food and beverages, construction, wholesale, hotels and restaurants, other business services & 55 & Forestry, tobacco, leather & 0,1 \\
	\hline
	\end{tabular}
}
\end{center}
\end{table}

\section*{Model 3: Inter-Sectoral Analysis}

In Model 3, one observes a strong right shift in the distribution of largest avalanche sizes (see table~\ref{table:LAI3}). In the scenario of high resilience, one important change is that Luxembourg no longer experienced smaller avalanche sizes compared to the rest of the countries; in this simulation the largest avalanche covered the entire economy. With updating in production after receiving a shock, all countries experienced larger avalanches. In particular, all countries, except Italy, experienced avalanches of size 59, covering the entire economy, triggered by several sectors. The largest avalanche triggered in Italy was of 57 sectors. 

The sectors that triggered the largest avalanches in all countries are still activities in the service sector such as Common triggers of the largest avalanches are other business services (20 countries), construction (18 countries), food and beverages (17 countries), wholesale (15 countries), and chemicals
(7 countries). The sectors that triggered the smallest avalanches remained similar; in this simulation we still found some activities of the primary sector and some services. In all countries the sector uranium and thorium ores triggered the smallest avalanche, metal ores in 11 countries, tobacco in 9 countries, recovered secondary raw materials in 9 countries, and water transport services in five.

In the low-resilience case, the sectors other business services, construction, and wholesale, among others, remained as the triggers of the largest avalanches in most countries. The sector uranium and thorium ores was the most common trigger of the smallest avalanches in all countries (see table~\ref{table:LAI3I}).

Finally, we highlight the fact that the financial sector was the sector that triggered the largest avalanches in many countries particularly when the capacity threshold was the lowest and the shock was the highest which is the worst scenario analyzed. This means that, in order for the financial sector to have a large impact in the countries, the economies have to be weak enough and the shock needs to be large. This gives a first insight on the scenario that led to the spread of the 2008 financial crisis.

\begin{table}[htbp]
\caption{Model 3, High Resilience: Sectors that triggered the largest and smallest avalanches in each country.}
\label{table:LAI3}
\begin{center}
\scalebox{0.75}{
	\begin{tabular}[c]{|c|p{7.5cm}|c|p{7.5cm}|c|p{8cm}|c|}
	\hline
	 & \multicolumn{2}{|c|}{\textbf{Largest Avalanche}} & \multicolumn{2}{|c|}{\textbf{Smallest Avalanche}}\\ \hline
	\textbf{Countries} & Triggering sector & Size & Triggering sector & Size\\ \hline
	Austria & Basic Metals, machinery and equipment, electrical energy, construction, wholesale, and other business services & 59 & Uranium and thorium ores, metal ores, private households with employed persons, and fishing & 0,4\\ \hline
	Belgium & Chemicals, wholesale, other business services,  food and beverages, and construction & 59,58 & Uranium and thorium ores, recovered secondary raw materials, private households with employed persons & 0,2\\ \hline
	Czech Republic & Food and beverages, rubber and plastic, basic metals, motor vehicles, construction, wholesale, and other business services & 59 & Private households with employed persons, fishing, uranium and thorium ores & 0,3\\ \hline
	Denmark & Agriculture, food and beverages, construction, wholesale, land transport, water transport, real-estate, other business services & 59 & Uranium and thorium ores, private households with employed persons, metal ores & 0,2\\ \hline
	Estonia & Food and beverages, wood, construction, wholesale, land transport, and other business services & 59 & Coal and lignite peat, uranium and thorium ores, tobacco, and private households with employed persons & 0\\ \hline
	Finland & Food and beverages, paper, chemicals, basic metals, radio television and communication equipment, construction, wholesale, and other business services & 59 & Uranium and thorium ores, private households with employed persons, and tobacco & 0,1\\ \hline
	France & Food and beverages, chemicals, construction, wholesale, other business services & 57 & Tobacco, private households with employed persons, coal and lignite peat, and metal ores & 1,2\\ \hline
	Germany & Chemicals, machinery and equipment, motor vehicles, other business services & 59 & Uranium and thorium ores, private households with employed persons, and fishing & 0,1\\ \hline
	Greece & Food and beverages, coke refined petroleum, construction, other business services & 59 & Uranium and thorium ores, private households with employed persons, office machinery, and recovered secondary raw materials & 0,2\\ \hline
	Hungary & Food and beverages, radio television and communication equipment, and other business services & 59 & Uranium and thorium ores, private households with employed persons, and fishing & 0,3\\ \hline
	Ireland & Construction, other business services, chemicals, food and beverages & 59,58,57 & Crude petroleum and natural gas, uranium and thorium ores, metal ores, furniture, and private households with employed persons & 0\\ \hline
	Italy & Other business services, construction, chemicals, food and beverages, and wholesale & 57,48,46,43 & Crude petroleum and natural gas, uranium and thorium ores, metal ores, furniture and private households with employed persons & 1\\ \hline
	Lithuania & Crude petroleum and natural gas, chemicals, food and beverages, electrical energy, and wholesale & 59,58 & Uranium and thorium ores, metal ores, tobacco, coke refined petroleum, and private households with employed persons & 0\\ \hline
	Luxembourg & financial intermediaries, services auxiliary to financial intermediaries, other business services, construction & 59,58,57 & Uranium and thorium ores, metal ores, tobacco, leather, recovered secondary raw materials, and private households with employed persons & 0\\ \hline
	Netherlands & Food and beverages, coke refined petroleum, construction, and other business services & 59 & Uranium and thorium ores, private households with employed persons, forestry & 0,4\\ \hline
	Poland & Agriculture, food and beverages, electrical energy, construction, wholesale, and other business services & 59 & Crude petroleum and natural gas, uranium and thorium ores, water transport & 0\\ \hline
	Portugal & Food and beverages, construction, other business services & 59 & Uranium and thorium ores, private households with employed persons, and metal ores & 0,1\\ \hline
	Romania & Agriculture, food and beverages, electrical energy, crude petroleum and natural gas, basic metals, construction, and wholesale & 59,57 & Uranium and thorium ores, membership organization services, recreational cultural and sporting services, other services, and private households with employed persons & 0\\ \hline
	Slovenia & Basic metals, construction, wholesale,and  other business services & 59 & Uranium and thorium ores, private households with employed persons, and tobacco & 0,2\\ \hline
	Spain & Construction, agriculture, food and beverages, wholesale, and other business services & 59,58 & Uranium and thorium ores, private households with employed persons, and tobacco & 0,1\\ \hline
	Sweden & Motor vehicles, wholesale, real-estate, and other business services & 59 & Uranium and thorium ores, other mining and quarrying products, tobacco, wearing apparel, and radio television and communication equipment & 0\\ \hline
	UK & Construction, wholesale, other business services, and real-estate & 59,58 & Uranium and thorium ores, metal ores, recovered secondary raw materials, and private households with employed persons & 0\\
	\hline
	\end{tabular}
}
\end{center}
\end{table}

\begin{table}[htbp]
\caption{Model 3, Low Resilience: Sectors that triggered the largest and smallest avalanches in each country.}
\label{table:LAI3I}
\begin{center}
\scalebox{0.75}{
	\begin{tabular}[c]{|c|p{5cm}|c|p{9cm}|c|c|c|}
	\hline
	 & \multicolumn{2}{|c|}{\textbf{Largest Avalanche}} & \multicolumn{2}{|c|}{\textbf{Smallest Avalanche}}\\ \hline
	\textbf{Countries} & Triggering sector & Size & Triggering sector & Size\\ \hline
	Austria & 37 sectors & 59 & Uranium and thorium ores, metal ores,  private households with employed persons, fishing & 0\\ \hline
	Belgium & 25 sectors & 59 & Uranium and thorium ores, recovered secondary raw materials, and private households with employed persons & 0\\ \hline
	Czech Republic & 34 sectors & 59 & Private household with employed persons, fishing, and uranium and thorium ores & 0,4\\ \hline
	Denmark & 37 sectors & 59 & Uranium and thorium ores, private households with employed persons, and metal ores & 0,2\\ \hline
	Estonia & 32 sectors & 59 & Coal and lignite peat, uranium and thorium ores, tobacco, and private households with employed persons & 0\\ \hline
	Finland & 33 sectors & 59 & Uranium and thorium ores, private households with employed persons, and tobacco & 0,1\\ \hline
	France & 38 sectors & 57 & Tobacco, metal ores, private households with employed persons, and coal and lignite peat & 5,12,13,15\\ \hline
	Germany & 31 sectors & 59 & Uranium and thorium ores, private households with employed persons, and fishing & 0,6\\ \hline
	Greece & 24 sectors & 59 & Uranium and thorium ores, private households with employed persons, and recovered secondary raw materials & 0,6\\ \hline
	Hungary & 26 sectors & 59 & Uranium and thorium ores, private households with employed persons, and fishing & 0,7\\ \hline
	Ireland & 14 sectors & 59 & Crude petroleum and natural gas, uranium and thorium ores, metal ores, furniture, and private households with employed persons & 0\\ \hline
	Italy & Food and beverages, chemicals, construction, wholesale, and other business services & 59 & Crude petroleum and natural gas, uranium and thorium ores, metal ores, furniture, and private households with employed persons & 1\\ \hline
	Lithuania & 15 sectors & 59 & Uranium and thorium ores, metal ores, tobacco, coke refined petroleum, and private households with employed persons & 0\\ \hline
	Luxembourg & Financial intermediaries, services auxiliary to financial intermediaries, and other business services  & 59 & Uranium and thorium ores, metal ores, tobacco, leather, recovered secondary raw materials, and private households with employed persons & 0\\ \hline
	Netherlands & 26 sectors & 59 & Uranium and thorium ores, private households with employed persons, and forestry & 0,8\\ \hline
	Poland & 30 sectors & 59 & Crude petroleum and natural gas, uranium and thorium ores, other mining and quarrying products, and water transport & 0,14\\ \hline
	Portugal & 32 sectors & 59 & Uranium and thorium ores, private households with employed persons, and metal ores & 0,12\\ \hline
	Romania & 23 sectors & 59 & Uranium and thorium ores, membership organization services, recreational cultural and sporting services, other services, and private households with employed persons & 0\\ \hline
	Slovenia & 30 sectors & 59 & Uranium and thorium ores, private households with employed persons, and tobacco & 0,3\\ \hline
	Spain & 22 sectors & 59 & Uranium and thorium ores, private households with employed persons, and tobacco & 0,15\\ \hline
	Sweden & 30 sectors & 59 & Uranium and thorium ores, other mining and quarrying products, tobacco, wearing apparel, and private households with employed persons & 0\\ \hline
	UK & 35 sectors & 59 & Uranium and thorium ores, metal ores, recovered secondary raw materials, and private households with employed persons & 0\\
	\hline
	\end{tabular}
}
\end{center}
\end{table}

 \end{document}